\documentclass{emulateapj}

\pdfoutput=1
\usepackage{epstopdf}
\usepackage{amsmath}
\usepackage{comment} 
\usepackage{appendix}
\usepackage{color}

\slugcomment{\today}
\newcommand{\xxScale}{1.15}  

\begin{document}
\title{Color Dispersion and Milky Way Reddening Among Type Ia Supernovae}


\shorttitle{The Color of Type Ia Supernovae}
\shortauthors{Scolnic et al.}

\author{
Daniel M. Scolnic,\altaffilmark{1}
Adam G. Riess,\altaffilmark{1}
Ryan J. Foley,\altaffilmark{2}
Armin Rest,\altaffilmark{3}
Steven A. Rodney,\altaffilmark{1}
Dillon J. Brout,\altaffilmark{1}
David O. Jones\altaffilmark{1}
}

\altaffiltext{1}{Department of Physics and Astronomy, Johns Hopkins University, 3400 North Charles Street, Baltimore, MD 21218, USA}
\altaffiltext{2}{Harvard-Smithsonian Center for Astrophysics, 60 Garden Street, Cambridge, MA 02138}
\altaffiltext{3}{Space Telescope Science Institute, 3700 San Martin Drive, Baltimore, MD 21218}

\begin{abstract}

Past analyses of Type Ia Supernovae (SNe\,Ia) have identified an irreducible scatter of $5-10\%$ in distance widely attributed to an intrinsic dispersion in luminosity.  Another, equally valid, source of this scatter is intrinsic dispersion in color.  Misidentification of the true source of this scatter can bias both the retrieved color-luminosity relation and cosmological parameter measurements.  The size of this bias depends on the magnitude of the intrinsic color dispersion relative to the distribution of colors that correlate with distance.  We produce a realistic simulation of a misattribution of intrinsic scatter, and find a negative bias in the recovered color-luminosity relation, $\beta$, of $\Delta \beta\approx -1.0$ ($\sim33\%$) and a positive bias in the equation of state parameter, $w$, of $\Delta w \approx +0.04$ ($\sim4\%$).  We re-analyze current published data sets with the assumptions that the distance scatter is predominantly the result of color.  Unlike previous analyses, we find that the data are consistent with a Milky Way reddening law ($R_V=3.1$), and that a Milky Way dust model better predicts the asymmetric color-luminosity trends than the conventional luminosity scatter hypothesis.  We also determine that accounting for color variation reduces the correlation between various Host galaxy properties and Hubble residuals by $\sim20\%$.  
\end{abstract}

\section{Introduction}
\label{sec:intro}

Since the initial discovery of evidence for cosmic acceleration (\citealp{Riess98} , \citealp{Saul99}), there has been a concerted effort to discover increasingly larger samples of Type Ia Supernovae (SN\,Ia) and also to probe the systematic uncertainties in the current samples.  SN\,Ia measurements are still the optimal method to measure the equation of state of dark energy $w=\rho c^2$ since SNe\,Ia can be observed in a redshift range where dark energy is dominant and because of their high precision.  In order to optimize the use of SNe\,Ia as standard candles to determine distances, the majority of the SN\,Ia light curve fitters (e.g MLCS2k2; \citealp*{Jh07}, SALT2; \citealp{2007A&A...466...11G}, SiFTo; \cite{2008ApJ...681..482C}, CMAGIC; \cite{2009ApJ...699L.139W}, SNooPY; \cite{SNooPy}, BAYESN; \cite{Mandel/etal:2011}) include two corrections to the observed peak magnitude of the SN: one using the width/slope of the light curve and the other using the color of the light curves.  The width/slope correction for various light curve fitters all account in some manner for the ``Phillips relation" \citep{phillips93}, be it the spectral adaptive light-curve template method (SALT2, SiFTO), the multicolor light-curve shape method (MLCS2k2), the color-magnitude intercept method (CMAGIC), a Bayesian hierarchical method (BayesN) or the $\Delta m_{15}(B)$ method itself (SNooPY).  

A more fundamental difference between these fitters is how they interpret heterogeneous SN\,Ia colors.  Some light curve fitters like SALT2 or SiFTo find an empirical relation between color and luminosity, called $\beta$ in SALT2, while MLCS2k2 assumes that the color-luminosity relation follows the Milky Way reddening law.  Regardless of the approach, the corrections reduce the dispersion in distance to $\sim5-10\%$, which is assumed to be intrinsic but may result from unmodeled effects (e.g. \citealp{kimgp}, \citealp{Chotard_2011}).  To ensure that all cosmology fits have reasonable parameter errors ($\chi^2_{\nu}\sim1$), an intrinsic scatter of $0.05-0.15$ mag is added in quadrature to the distance modulus uncertainties. 

 Historically, this irreducible scatter in the distance modulus residuals relative to a best fit cosmology, also called `Hubble residual scatter', has been attributed to random, achromatic variations in the luminosity.  This `luminosity variation' is the variation in SN\,Ia brightness that does not correlate with distance and is independent of the light curve corrections.  While it is relatively simple to apply the assumption that Hubble residual scatter is due to luminosity variation, there is little, if any, direct evidence to support this claim.  Variations of color and shape in the SN light curves, which do not correlate with distance in the way expected by light curve fitters, present equally plausible alternatives.  We denote color variation as the variation in colors for a fixed distance, which can be observed as the variation in distances at a fixed color.  \cite{Mar11} (hereafter M11) presents a mathematical formalism for assigning this ``intrinsic scatter" of SN\,Ia to either the luminosity, color or stretch variation or a combination of the three.  Here, we call this scatter the `residual scatter', or $\sigma_r$, as it is needed to explain the scatter in the Hubble residuals.  \cite{kessler_13_var} (hereafter K13) explores various sources of this residual scatter and shows that misidentifying the source of scatter biases the recovery of $\beta$ by up to $10\%$ and $w$ up to $5\%$ when assuming a broad, fixed color distribution.  We revisit this assumption here. 

There is evidence to suggest that color variation is non-negligible.  \cite{Fo11} and \cite{Chotard_2011} both examine the relations between Silicon and Calcium features of SN\,Ia spectra and SN\,Ia color and conclude there must be color variation on the order of what is needed to explain the Hubble residual scatter.  \cite{Jh07} analyzed $+35$ day nebular colors \citep{Lira}, a phase when light curve shape dependent color variation is minimized, and found there must be a similarly high amount of color variation.  There is also preliminary evidence that color variation may partly account for the trends between Hubble residuals and host galaxy properties \citep{Ch13}.

In this paper, we use publicly available data from SDSS \citep{Ho08}, SNLS3 \citep{2007A&A...466...11G} and nearby samples (see \citealp{Conley_etal_2011} for a review) and the SNANA simulator \citep{K09b} and explore how misattribution of the source of residual scatter and ignorance of the underlying color distribution affect $\beta$ and $w$ estimation.  In \S\ref{sec:lcfits} we present an analysis of the different components of the observed color and explain how the bias in the SN color-luminosity relation depends on both the source of residual scatter and the underlying distribution of color.  In \S\ref{sec:model} we show that if color variation causes the residual scatter, the empirical relation between SN\,Ia distance and color is well represented by a Milky Way Reddening law.  In \S\ref{sec:consequences} we discuss implications of different models for SN\,Ia color, including effects on $w$ recovery, how the distance residual bias introduces what appears to be $\beta$ evolution and how host-galaxy-luminosity correlations can be partly explained by the bias.  Our discussion and conclusions are in \S\ref{sec:discuss} and \S\ref{sec:conclus}.


\section{The dependance of the color-luminosity relation on the source of scatter}
\label{sec:lcfits}
\subsection{The Different Sources of Residual Scatter}
\nobreak

In order to understand the color-luminosity relation of SN\,Ia, we must define the different components of SN\,Ia color and how each component is treated by light curve fitters to determine distances.  For most of the analysis in this paper, we employ the SALT2 light curve fitter as its empirical framework easily allows for different assumptions and it is one of if the most widely used light curve fitters.  

The distance modulus $\mu$ determined by SALT2 for each SN\,Ia is expressed as 
\begin{equation}
\mu =m_B - M_0 +\alpha x_1 - \beta c
 \label{eqn:salt2}
\end{equation}
where $m_B$, $x_1$ and $c$ are the individual fit parameters representing the rest frame B band peak brightness, stretch of the light curve, and color of the SN respectively.  $M_0$, $\alpha$ and $\beta$ are parameters that represent the absolute magnitude of a standard Ia, the slope of the stretch-luminosity relation and the slope of the color-luminosity relation, respectively.  
We follow M11 since it accounts for residual scatter in any of the SALT2 fit parameters and has the advantage of separating the determination of the SALT2 nuisance parameters from a specific cosmology (see Appendix).  The error of $\mu$ is assumed to be the quadrature sum of the ``noise", $\sigma_n$, and residual scatter applicable to the model,$\sigma_r$, such that $\sigma^2_{tot}=\sigma^2_n+\sigma^2_r$.

The observed color, $c_{\textrm{obs}}$, can be expressed as
\begin{equation}
c_{\textrm{obs}}=c_{\textrm{mod}}+c_r+c_n~,
\end{equation}
where $c_{\textrm{mod}}$ is the model color which is the component of SN color that is linearly correlated with luminosity by $\beta$.  The residual color $c_{r}$ is the random color component uncorrelated with luminosity, and $c_{n}$ is the noise of the color measurement. Conventionally, $c_r\equiv0$ and the total residual scatter $\sigma_{\textrm{r}}$ is given as a single number that represents only the residual scatter in the peak B-band luminosity of SN\,Ia, $m_B$, as it correlates directly with $m_b$ or $M_0$.  M11 addresses this assumption, and allows the residual scatter to represent the residual scatter in $m_B$, $c$, $x_1$ or combinations of the three.  More generally, this residual scatter for each SN is:
\begin{multline}
\sigma_{\rm{r}}^2 = \displaystyle \sigma^2_{{{m}_r}} + \alpha^2\sigma^2_{{{x_1}_r}}+\beta^2 \sigma^2_{{c_r}}
+2\alpha\Sigma_{{{mx_1}_r}} \\
-2\beta\Sigma_{{{mc}_r}}-2\alpha\beta\Sigma_{{{x_1c}_r}} .
\label{eqn:sigmai}
\end{multline}
$\Sigma$ represents the $3\times3$ residual scatter matrix in $m_B, x_1,$ and $c$ and $\sigma^2_{m_r},\sigma^2_{{x_1}_r},\sigma^2_{c_r}$ are its diagonal components.   It is important to note that since $\sigma_r$ includes $\alpha$ and $\beta$ terms, these coefficients play a role in not only correcting the distances but also propagating the uncertainty of each distance.  If $\alpha$ and $\beta$ were not included in the uncertainty, we would find significantly lower values of $\alpha$ and $\beta$.

\begin{figure}
\centering
\epsscale{\xxScale}  
\plotone{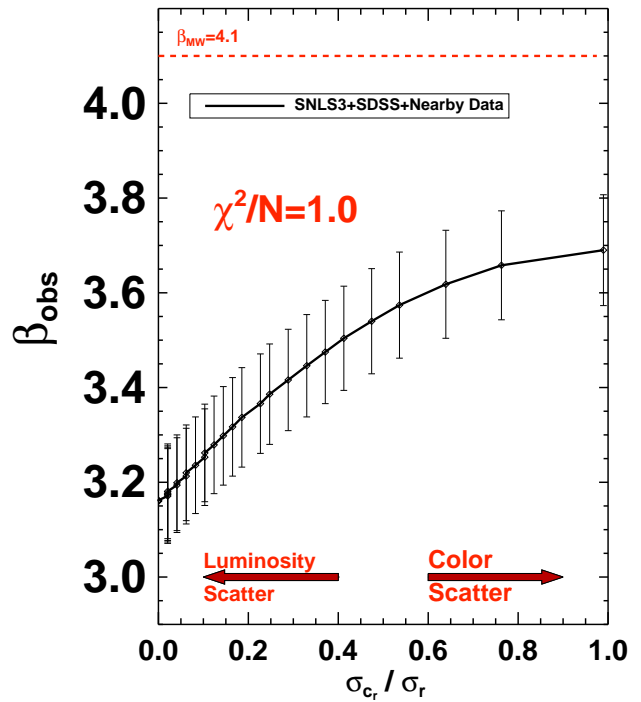}
\caption{From real data, the effects of different assumptions for the source of the residual intrinsic scatter $\sigma_{r}$ on the observed $\beta$ parameters.  The lower x-axis shows the relative weight of the residual color scatter ($\sigma_{c_r}$) relative to the full residual scatter ($\sigma_{r}$).  For all assumptions, the total residual scatter has a magnitude such that the reduced $\chi^2$ of the Hubble residuals is unity.  For plotting only, $\sigma_{c_r}$ is weighted such that $\sigma_{c_r}^2=\sigma_{r}^2-\sigma_{m_r}^2$. }
\label{fig:system}
\end{figure}

To understand the consequences of incorrectly attributing the source of the observed distance scatter, we analyze publicly available data from SDSS, SNLS3 and Nearby samples.  We include 91 SDSS SNe\,Ia \citep{Ho08} and 241 SNLS3 SNe\,Ia \citep{Gu10} and a Nearby sample comprised of 186 SNe\,Ia from a variety of sources, most of which are described in \citep{Conley_etal_2011}.  The only additions to the \cite{Conley_etal_2011} set are 67 SNe\,Ia from the CfA4 sample \citep{2009ApJ...700..331H} and 34 additional CSP SNe\,Ia \citep{2010AJ....139..519C}.  To fit the light curves, we use the SNANA SALT2 light curve fitter and its provided files for defining the filter transmission functions.

\begin{deluxetable}{lcccccc}[!H]
\tablecaption{Underlying Color Populations
\label{tab:evot}}
\tablehead{
\colhead{Survey~~} &
\colhead{$\sigma_{c_{\textrm{obs}}}~~$} &
\colhead{$\sigma_{c_n}~~$} &
\colhead{Scatter~~} &
\colhead{Components~~} &
\colhead{ $\sigma_{c_{\textrm{mod}}}$~~}  \\
~ & ~ & ~ & ~ & $\sigma_{m_r}, \sigma_{c_r}$ &~ \\
~ & [Obs.] & [Obs.] & [Assumed] & [Assumed] &[Derived] \\}
\startdata
SNLS3 & 0.087&0.043&Lum. & $[0.14,0.0]$ & 0.076 \\
SDSS & 0.076&0.039&Lum. & $[0.14,0.0]$ & 0.065\\
Nearby & 0.094&0.032&Lum. & $[0.14,0.0]$ & 0.088  \\

SNLS3 & 0.087&0.043&Color & $[0.0,0.04]$ & 0.064 \\
SDSS & 0.076&0.039&Color & $[0.0,0.04]$ & 0.051\\
Nearby & 0.094&0.032&Color & $[0.0,0.04]$ & 0.079 \\

\enddata
\tablecomments{The different components of the observed color distribution given various assumptions of the residual scatter model.  $\sigma_{c_{\textrm{obs}}}$ and $\sigma_{c_n}$ are found from the real data.  The scatter source is assumed and $\sigma_{c_{\textrm{mod}}}=\sqrt{\sigma^2_{c_{\textrm{obs}}}-\sigma^2_{c_n}-\sigma^2_{c_r}}$. }
\end{deluxetable}

  In Figure 1, we show the dependence of $\beta$ from the full data set on the fraction of the residual scatter assumed to result from color.  We ignore stretch here as luminosity and stretch variation affect the color-luminosity relation in a similar manner.   We find the higher the contribution of color to the residual scatter component, the higher the value of $\beta$ recovered.  Because past analyses have assumed $\sigma_{c_r}=0$, they found the lowest possible value of $\beta$.  Attributing the residual scatter to color changes the retrieved value of $\beta$ from 3.2 to 3.7.  Interestingly, we find that $\beta$ is relatively close to the Milky Way extinction law of $\beta=4.1$ when the residual scatter is entirely attributed to color variation, a plausible but unproven possibility.  In Figure 1, the SNLS3, SDSS and Nearby samples are combined, though we also find that the dependance of $\beta$ on the source of the residual scatter to be somewhat different for each survey.  When residual scatter is attributed entirely to color,  we find $\beta=3.80\pm0.161$, $\beta=3.65\pm0.124$ and $\beta=3.20\pm0.102$ for the Nearby, SNLS3 and SDSS samples respectively. For the Nearby sample, the value of $\beta$ is consistent with the range of Milky Way extinction.  When residual scatter is attributed entirely to luminosity,  we find $\beta=3.34\pm0.171$, $\beta=3.02\pm0.164$ and $\beta=2.91\pm0.210$ for the Nearby, SNLS3 and SDSS samples respectively.  An explanation for these differences will be presented in the following section.


\subsection{Knowledge of the Color Distribution}

To quantify the bias in the recovery of $\beta$ when residual color scatter is ignored, we must understand the consequences of the previous assumption that $\sigma_{c_r}=0$ in Eq. 3.  We define $\beta_{\textrm{mod}}$ as the color-luminosity relation if there is no color noise or color variation ($c_{\textrm{obs}}=c_{\textrm{mod}}$), and $\beta_{\textrm{obs}}$ as the color-luminosity relation if there is non-zero color noise and/or color variation ($\sigma_{c_r},\sigma_{c_n}\ne0$).  For a distribution of modeled colors defined by a gaussian of width $\sigma_{c_{\textrm{mod}}}$, we expect the bias in beta recovery to be (see Appendix B for review):
\begin{equation}
\frac{\beta^2_{\textrm{obs}}}{\beta^2_{\textrm{mod}}}\approx \frac {\sigma^2_{c_{\textrm{mod}}}}{\sigma^2_{c_{\textrm{mod}}}+\sigma^2_{c_n}+\sigma^2_{c_r}}~.
\end{equation}

If $ ({\sigma^2_{c_n} + \sigma^2_{c_r} }) \ll \sigma^2_{c_{\textrm{mod}}}$, then $\beta_{\textrm{obs}} \approx \beta_{\textrm{mod}}$ as assumed in past analyses.  However, if $({\sigma^2_{c_n} + \sigma^2_{c_r} })  \sim \sigma^2_{c_{\textrm{mod}}}$, then $\beta_{\textrm{obs}} < \beta_{\textrm{mod}}$. Therefore, the change in $\beta_{\textrm{obs}}$ from the true color-luminosity correlation $\beta_{\textrm{mod}}$ is dependent not only on the presence of residual color scatter but also its size in comparison to the underlying color distribution of SN\,Ia, $\sigma_{c_{\textrm{mod}}}$.

In order to find the bias in $\beta$ and verify Eqn. 4, we simulate SN\,Ia samples with different magnitudes of residual color scatter ($\sigma_{c_r}$) and widths of gaussian, color model distributions ($\sigma_{c_{\textrm{mod}}}$).  Any simulation must replicate the observed color distribution of the real data $\sigma_{c_{\textrm{obs}}}$.  Because $\sigma^2_{c_{\textrm{obs}}}=\sigma^2_{c_{\textrm{mod}}}+\sigma^2_{c_n}+\sigma^2_{c_r}$, increasing the magnitude of $\sigma^2_{c_r}$ in the simulations requires that $\sigma^2_{c_{\textrm{mod}}}$ is decreased.  The main divergence between this analysis and that of K13 is that K13 does not change $\sigma_{c_{\textrm{mod}}}$ when they vary $\sigma_{c_r}$.  In Table 1, we show $\sigma_{c_{\textrm{obs}}}$ and $\sigma_{c_n}$ for each sample, and after assuming the magnitude of color scatter $\sigma_{c_r}$, we find $\sigma_{c_{\textrm{mod}}}$.  

For the simulations, we use the SNANA \citep{K09b} simulator, which allows a user to incorporate information such as actual weather history, PSF characteristics, spectroscopic follow-up strategies and underlying distributions of color and stretch all towards mimicking a true supernova survey.  In order to simulate color or luminosity scatter, we follow the SNANA procedure that adds random magnitude offsets generated for each SN and observed filter to each light-curve point measured in that filter.  This process is called `color smearing'  and for simulating luminosity scatter, the additional, random magnitude offset is the same for all filters, while for color scatter, the additional, random magnitude offset is different for each filter.


To estimate the largest bias possible in the recovery in $\beta$, we simulate supernova samples with scatter entirely due to residual color variation ($\sigma_{c_r}=0.04$) but in the recovery of $\beta$ we assume $\sigma_{c_r}=0$ and therefore misattribute all of the scatter to luminosity variation ($\sigma_{m_r}$).  The simulations used for this exercise have the characteristics of the SNLS3 survey (e.g. weather, cadence, seeing, $\sigma_{c_n}\approx0.035$), and we fix the simulation input $\beta=4.1$ so that the color-luminosity relation is consistent with extinction in the Milky Way.  We show, in Figure 2, how the magnitude of the bias in the recovery of $\beta_{\textrm{obs}}$ depends on the relative size of the simulated color variation $\sigma_{c_r}$ to the width of the color distribution $\sigma_{c_{\textrm{mod}}}$.  The trend seen from the simulations is in decent agreement with what is predicted from Eq.~4 and discrepancies are likely due to covariances between the light curve fit parameters.  While we chose to input $\beta=4.1$ in the simulation, the trend seen in Fig. 2 would be similar for different input $\beta$ values.  

To best estimate this bias in $\beta_{\textrm{obs}}$ for the individual SDSS, SNLS3 and Nearby samples, we take the $\sigma_{c_{\textrm{mod}}}$ value for each sample in the Color-Only case in Table 1.  We find from Fig. 2 that for the Color-Only case we recover on average $\beta_{\textrm{obs}} \approx3.1$ for these samples, $1$ lower than the input value and consistent with the value of $\beta$ seen in the literature \citep{Conley_etal_2011}.  Therefore, we find that the value of $\beta$ regularly quoted as disproving the hypothesis that colors of SNe\,Ia follow a Milky Way Reddening law falls out naturally from a simulation that has two simple assumptions: the true color-luminosity relation follows the Milky Way Reddening law and Hubble residual scatter is due to color variation but misattributed to luminosity variation.

 We also offer an explanation of why there should be disagreement on $\beta$ between these supernova samples.  Since the $c_{\textrm{obs}}$ distributions of these three samples are different, then for the same $\sigma_{c_r}$ value we would expect that $\sigma_{c_{\textrm{mod}}}$ is different for each sample, and therefore $\beta_{\textrm{obs}}$ should vary.  This claim is reasonable as the Nearby sample should contain more SNe\,Ia with higher extinction values than in the SDSS or SNLS3 samples and should have a higher $\sigma_{c_{\textrm{mod}}}$ value.  

\begin{figure}
\centering
\epsscale{\xxScale}  
\plotone{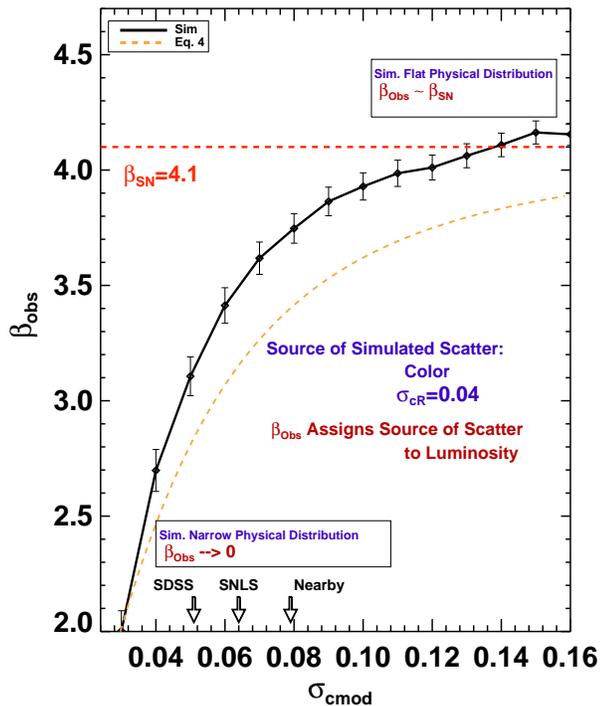}
\caption{The dependance of $\beta_{\textrm{obs}}$ the width of the true color distribution, $\sigma_{c_{\textrm{mod}}}$ from simulations with residual scatter due entirely to color variation and $\beta_{\textrm{SN}}=4.1$.  In finding $\beta_{\textrm{obs}}$, residual scatter is misattributed to luminosity. Each data point represents its own simulation. We mark on the x-axis the likely value of $\sigma_{c_{\textrm{mod}}}$ for each sample from Table 1.}
\label{fig:example}
\end{figure}

\section{Two Degenerate Models of Supernova Color?}
\label{sec:model}
\subsection{A Physical Color Model}
In the previous section, we showed that if $\beta=4.1$ and the residual scatter is due to color variation but misattributed to luminosity, then an analysis would find $\beta \approx3.1$ in agreement with past studies.  Now we explore the physical assumption that the model color may solely be due to reddening.  In this approach, the observed color, $c_{\textrm{obs}}$, can be expressed as
\begin{equation}
c_{\textrm{obs}}=c_{\textrm{dust}}+c_r+c_n,
\end{equation}
where the equation is of the same form as Eq. 2 but $c_{\textrm{mod}}$ is now replaced with $c_{\textrm{dust}}$.  MLCS2k2, which is founded on astrophysical assumptions, denotes an unreddenned color to have an $A_V=0$, that according to  \cite{K09a}, is roughly analogous to a $c\approx -0.10$ for SALT2\footnote{From Fig. 8 in \cite{2009ApJ...700.1097H}, $c$ appears to be $\approx-0.05$ for $A_V=0$, though there is a significant amount of scatter.}.  For this physical model of color, we would therefore expect $c_{\textrm{dust}}\ge-0.10$ and that residual color scatter explains colors bluer than $c_{\textrm{obs}}=-0.10$.  

The assumption that $c_{\textrm{mod}}$ is solely due to reddening has not only physical but also empirical motivations.  K13 found that the model distribution of color is best described by an asymmetric gaussian (see Appendix B for explanation) with a blue-ward standard deviation which is significantly shorter than the red-ward standard deviation.  A smaller blue range likely implies that most of the color of a SN is due to reddening, rather than some other color related property of the SN.  For the 306 SNe\,Ia in the combined SDSS+SNLS3+Nearby sample with small statistical color errors ($\sigma_c<0.04$), only 35 SNe have colors bluer than the $A_V$ cutoff of $c=-0.1$ ($\sim 11\%$).  A residual scatter of $\sigma_{c_r}=0.04$ is large enough to replicate in a simulation the blue side of the observed distributions of color given an input cutoff at $c=-0.1$.

\begin{figure}
\centering
\epsscale{\xxScale}  
\plotone{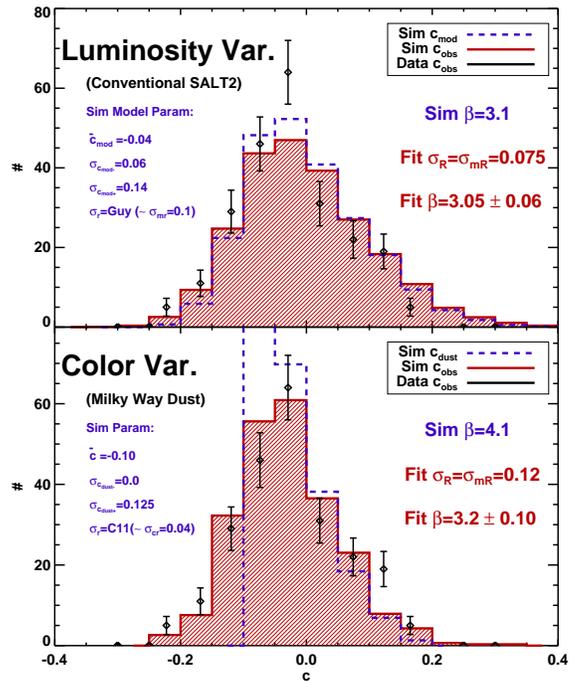}
\caption{The $c_{\textrm{obs}}$ distribution for SNLS3 simulations with the SALT2 luminosity-variation model and the `Milky Way' color-variation model.  The input ($c_{\textrm{mod}}$) distribution and observed distribution are shown for each simulation, as well as the true SNLS3 observed color distribution.  The parameters of the input distribution for each simulation are given.}
\label{fig:MLCS}
\end{figure}

 \begin{figure}
\centering
\epsscale{\xxScale}  
\plotone{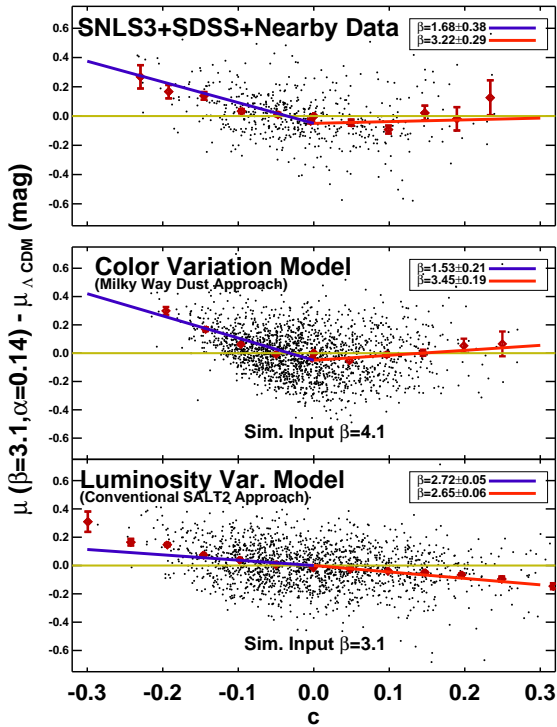}
\caption{The relation between Hubble residuals and color for the full SNLS3+SDSS+Nearby sample, a simulated sample based on Milky Way-like extinction, and the conventional empirical model.  The parameters of the two simulations are shown in Fig. 4.  The slopes of the trend of Hubble residuals with colors for both blue ($c<0$) and red ($c>0$) colors are shown. }
\label{fig:hatfield}
\end{figure}

We wish to compare how the output color distributions from simulations of two significantly different models of color match the data.  For the first model, color is due to dust and random variation; in the second model we take the conventional SALT2 approach that there is no color variation.  For the Color Variation model, we create a simulation with $\beta_{\textrm{mod}}=4.1$, a model distribution that follows a reddening-only one-sided gaussian ($c_{\textrm{dust-min}}=-0.1, \sigma_{\textrm{{cdust+}}}=0.12$), and residual scatter due to color variation.  The standard deviation of the one-sided gaussian is set so that the output color distribution of this simulation best matches the data, and is roughly equivalent to $2\times \sigma_{c_{\textrm{mod}}}$ from Table 1.  For the Luminosity Variation model,  we create a simulation with $\beta_{\textrm{mod}}=3.1$ and residual scatter due to luminosity variation and the $c_{\textrm{mod}}$ distribution is taken from the default SNANA distribution described in K13.  

In the previous section, we simulated samples using the M11 `color smearing' method which added magnitude variations to each filter.  To allow for easier reproducibility of our analysis, we follow the K13 method in which $\sigma_r$ depends on wavelength.  For the Luminosity variation case, we take the SALT2 scatter model (hereafter called Guy10) which claims that scatter is relatively independent of wavelength and therefore that the residual scatter is dominated by luminosity.  For the Color variation case, we follow \cite{Chotard_2011} which presents two different color-dominated scatter models in which the scatter has a strong wavelength dependence.  Of the two Chotard models, we find that the model denoted as C11\_0 in K13 is better at reproducing the data.

In Figure~3, we present the color distributions of our two models: the luminosity-variation Guy10 model that is typically used in SALT2, and the color-variation dominated C11 model in which model color is solely due to reddening, called the Milky-Way model.  We find that the observed color distributions of both models reproduce the data.  For the Milky-Way model, when residual scatter is attributed to luminosity, we find a $\beta$ value near $3.1$ and a similar intrinsic dispersion value as seen in the data of $\sigma_{m_r}=0.12$.  This result is interesting because we find that a simulation with the physical, Milky Way reddening model of color yields the same global fit parameters as the real data itself and this physical model of color has one less free parameter and is thus favored by Occam's Razor.  The result is in agreement with the results of Fig. 2, though in this case we have an asymmetric $c_{\textrm{mod}}$ that is tied to a physical understanding of color.

While we find simulations with a Milky-Way reddening model can produce the same color distribution as the conventional SALT2 model, we ultimately wish to resolve which model is more accurate.  So far we have observed that the color variation model with $\beta=4.1$ and luminosity variation with $\beta=3.1$ are degenerate as they both yield $\beta$ values of $\approx3.1$ when scatter is attributed to luminosity.  We hope to break this degeneracy.  In Fig.~3, we observe that the Milky-Way model assumes that blue-ward of $c=-0.1$, the color results from noise and residual color scatter.  These bluer colors would not be expected to correlate with luminosity in the same way as the redder colors.  The conventional SALT2 luminosity-variation model assumes that blue and red colors would be identically correlated with distance.  

To test these two predictions, we analyze the Hubble residuals after including a color-correction with $\beta=3.1$ for all our samples.  In Fig.~\ref{fig:hatfield}, we show that in the SNLS3+SDSS+Nearby data, there are effectively different color-luminosity relations for $c>0$ and $c<0$. The conventional SALT2 model predicts that $\beta$ should be unchanged over the color range, while the MLCS2k2 Milky-Way model predicts a bifurcation because of the $A_V=0$ cutoff and the long exponential reddening tail.  The color-luminosity relations shown are found using simple, linear fits to the data.  We see that the bifurcated slopes of the real data appear to match those from the Milky-Way prediction ($<1\sigma$ differences) significantly better than the conventional SALT2 prediction ($2-3\sigma$ differences).  By analyzing each color range independently, we may break the degeneracy between the physical model of color and the conventional SALT2 model, and find that the physical model is empirically optimal.  The inconsistent $\beta$ values found for blue and red colors are also seen in \cite{suzuki_2012_aa}; they find a $\Delta \beta =1.48 \pm 0.36$ between SNe\,Ia with $c>0.05$ and $c<0.05$.  Lastly, we note that while the C11 color model is used for this test, the simple color-smearing model in SNANA produces similar results.

\subsection{A Bayesian Approach for Analyzing Supernova Color}
So far when we have compared the $\beta$ values from these two approaches towards SN\,Ia color, we have assumed, correct or otherwise, that the Hubble scatter from each model is due to luminosity.  We have done this in order to explore the biases that would be present in other SN\,Ia analyses, if that conventional assumption is correct.  We may also attempt to analyze the data when we assume that the Hubble residual scatter is entirely due to color variation and that there is a dust cutoff of $c>-0.1$.  Unfortunately, there is no formalism to incorporate any kind of Bayesian prior in SALT2.  We introduce here a simple Bayesian algorithm applied to SALT2 (hereafter called BALT) that allows for the possibility that color follows the physical model outlined above. Once SALT2 finds a color from the light curve fit, we apply a Bayesian prior \citep{riess96} to the color such that
\begin{equation}
c_{\textrm{B}}=\frac{1}{P} \int_{c>\bar{c}} c e^{-(c-c_{\textrm{obs}})/2\sigma_{c_n}^2}e^{-(c-\bar{c})^2/\tau_S(z)^2} \partial c.
\end{equation}
where $c_{\textrm{B}}$ is the corrected color, $c_{\textrm{obs}}$ is the color from the light curve fit, $c_n$ is the noise from the color measurement and $P$ is a normalization constant.  The second part of Eq. 6 describes the Bayesian prior for the model color as shown in the bottom panel of Fig. 3 where $\bar{c}$ is the blue cutoff of the distribution ($A_V=0$).  ${\tau_S(z)}$ describes the shape of the one sided gaussian due to extinction for a given redshift $z$ for each survey $S$; the dependance of $\tau$ on survey and redshift allows selection effects to be modeled (following \citealp{K09a}).  From the previous subsection, we expect that $\bar{c}=-0.1$ and $\tau(z=0)=\sigma_{c_\textrm{dust}}=0.11$.  At higher redshifts, $\tau$ decreases since only SNe with bluer colors are discovered and followed-up.  $\tau(z>0)$ may be determined from simulations with an input $\tau(z=0)$\footnote{For $\boldsymbol{\vec{z}}=[0.0,0.2,0.4,0.6,0.8,1.0]$, we find: $\boldsymbol{\tau_{\textrm{SDSS}}}=[0.11,0.085,0.055,-,-,-]$ and $\boldsymbol{\tau_{\textrm{SNLS3}}}=[0.11,0.11,0.11,0.105,0.085,0.07].$}.  We estimate the uncertainty of $\sigma_{c_\textrm{dust}}$ by varying this value for the model distributions in the simulations, and observing how well the simulated $c_{\textrm{obs}}$ distribution compares to the data.  Doing so, we find $\tau(z=0)=0.11\pm0.02$.

If Eq. 6 is applied to each SN color of the SDSS+SNLS3+Nearby sample, we find a very significant reduction in the total $\chi^2$ of the sample and the intrinsic dispersion needed to bring $\chi_\nu^2$ to unity.  Following the conventional SALT2 approach where scatter is attributed to luminosity, we determine that the total $\chi^2/N=801.6/518$ and $\sigma_r=0.09$.  With the BALT approach, setting $\beta=4.1$, the total $\chi^2/N=592.3/519$ and $\sigma_r=0.05$.  Interestingly, if we exclude the Nearby Sample, the total $\chi_\nu^2=1.01$ for the BALT approach whereas $\chi_\nu^2=1.56$ for the conventional SALT2 approach.  Part of the reason that the total intrinsic dispersion ($\sim0.0$ mag) is so low after the BALT correction for the SDSS and SNLS3 samples is that due to selection effects, the model color range of the SNe that are followed-up is very narrow.  We note that for this sample, simply forcing all SNe with $c<-0.1$ to have $c=-0.1$ reduces the $\chi_\nu^2$ to 1.26.   

The main argument against correcting the colors a posteriori is that it ignores covariances between color and the other fit parameters; however, from Guy10 we expect those covariances to be small.  There is currently work being done on a sophisticated Bayesian hierarchical approach to SALT2 (March et al. 2012), and the BALT algorithm applied here shows the promise of this approach.  The conventional method of not applying a prior to the color distribution is equivalent to applying a flat Bayesian prior, which itself may bias the analysis.  A more complete exploration of the BALT method will be presented in the upcoming PS1 Systematics paper (Scolnic et al. in prep).

\section{Consequences of Different Color Models}
\label{sec:consequences}

\subsection{Cosmological Implications}

So far we have shown that two prominent approaches for handling SN color are degenerate, and we have introduced a method to break this degeneracy.  Now we ask: to what degree do we bias our measurement of $w$ when we make an incorrect assumption about the nature of SN color and the source of SN scatter? 

\begin{figure}
\centering
\epsscale{\xxScale}  
\plotone{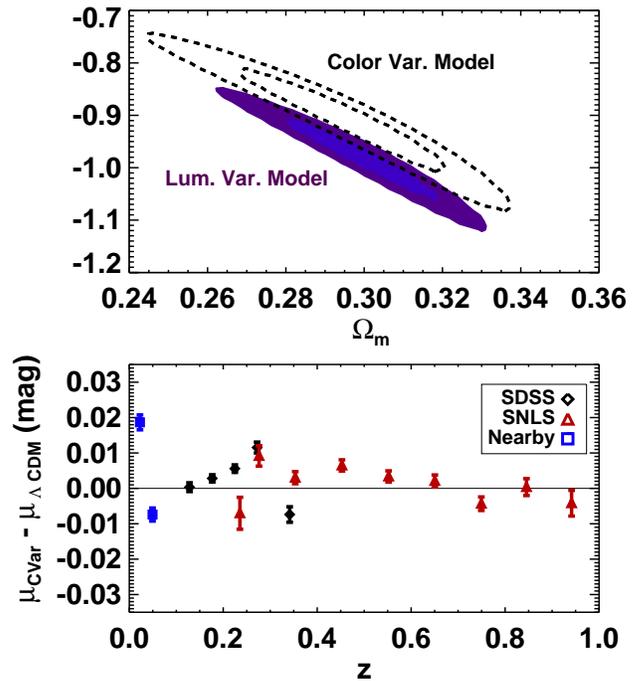}
\caption{(Top) Statistical constraints for the $68\%$ and $95\%$ confidence levels on $(\Omega_m,~w)$ from simulations using the conventional luminosity variation model and the Milky Way Dust + color variation model, including priors from CMB and BAO observations.  In both cases, the analysis assumes that residual scatter is due entirely to luminosity variation.  We assume a flat universe and constant dark energy equation of state.  (Bottom) The difference between observed distance and simulated distance for simulations based on the Milky Way Dust + color model ($\beta=4.1,\sigma_{c_r}=0.04$).  The observed distances have been derived by making the assumption that residual scatter is due to luminosity variation ($\beta=3.1,\sigma_{m_r}=0.11$).  The distances include a Malmquist correction. }
\label{fig:example}
\end{figure} 

To address this question, we use the SNANA program and its default simulation inputs to simulate the Nearby, SDSS and SNLS3 sample (see K13 for review).  We simulate two types of samples: one with the color variation model and the other with the luminosity variation model as described in the previous section.  For each survey, we simulate 10 samples with 5,000 SNe\,Ia so as to remove any statistical fluctuations between the simulated samples.  We then analyze all simulations attributing Hubble residual scatter to luminosity variation.  Following K13, these analyses include an empirically determined Malmquist correction, which corrects for selection effects and any biases introduced by the light curve fitter.

In Fig.~5 (bottom), we show the bias in the distance modulus for the simulation based on the color variation model for each survey.  We find that the bias is up to $\sim0.02$ mag for the Nearby sample, $\sim0.01$ mag for the SDSS sample and $\sim0.005$ mag for the SNLS3 sample.  Translating the bias in distances into an effect on retrieved cosmology depends on the priors used.  K13 uses priors from both WMAP \citep{Komatsu08} and SDSS-BAO \citep{Eisen05}, and we follow that method here, though we remark that the overlap between the statistical contours from these priors and that from different SNe analyses may hide inconsistencies between the SNe constraints.  In Fig.~5 (top), we present statistical cosmology constraints from two of our different simulations of a combined SNLS3+SDSS+Nearby Sample (5,000 SNe in each sample).  
\begin{deluxetable}{lll}
\tablecaption{Biases in $\beta$ and $w$ from misattributing the source of scatter
\label{tab:wtable}}
\tablehead{
\colhead{Sample} &
\colhead{$\Delta \beta$} &
\colhead{$\Delta w$}  \\
}
\startdata
SNLS3+SDSS+Nearby & $-0.92$ & $+0.037$ \\
SNLS3 only & $-0.90$ &$+0.042$ \\
SDSS only & $-1.05$ & $+0.023$ \\
\enddata
\tablecomments{The biases in $\beta$ and $w$ due to misattributing the source of scatter to luminosity variation for simulations with color-variation.  $w$ is found after including priors from SDSS-BAO and WMAP.}
\end{deluxetable}

These simulations reflect that over our multiple large simulations we find an average bias in $w$ for the color-variation simulation of $\approx+0.037$ for the full combined sample.  The results for the SDSS and SNLS3 samples individually are shown in Table 2.  

While the biases in $\beta$ are significantly larger than that in K13, the biases in $w$ found here and in K13 ($\Delta w\approx1-2\%$) are more similar.  The relative agreement between these studies is expected as both are probing the degeneracy of various color models.  The reason that the bias found here may be up to $2\times$ larger, though still small, is likely due to the the asymmetry in the color distribution.  

We also may compare the difference in $w$ when we apply the BALT color method to when we apply the conventional SALT2 luminosity method.  From the combined real SNLS3+SDSS+Nearby sample, when we attribute the residual scatter to luminosity variation, we find $w=-0.943\pm0.056$.  When we apply the BALT method, we find $w=-0.995\pm0.049$.  The difference of $\Delta w=+0.052$ is similar to that predicted from our simulations for the full sample.
\subsection{Host Galaxy Properties}
\label{sec:hosgal}
Since we have seen that the biases in $\beta$ and SN distances depend on the width of the color distributions, we now ask whether modest correlations between various host galaxy properties and Hubble residuals (e.g. \citealp{2010ApJ...715..743K},  \citealp{2010MNRAS.406..782S}, \citealp{gupta2011}) may be a result of these biases.  This question is further motivated by recent findings of correlations between host galaxy properties and SN colors \citep{Ch13}.

To explore this question, we find the widths of the color distributions for SNe in both high and low sSFR hosts, using sSFR values from \cite{2010MNRAS.406..782S}  (hereafter MS10).  We analyze the sSFR property, rather than mass, for this exercise as there is a clearer difference in the observed SN color distribution of subsamples split by sSFR than than by mass.  Mass subsamples appear to have different average colors, and we do not yet have the formalism to account for this fact.  Following  \S2.2, we derive the width of the model color distribution for the high sSFR hosts in SNLS3 to be $\sigma_{c_{\textrm{mod}}}=0.065$ mag and the width for the low sSFR hosts to be $\sigma_{c_{\textrm{mod}}}=0.050$ mag.  From Fig. 2, we extrapolate that the difference in $\beta$ from these two samples should be $\Delta \beta \approx-0.40$, roughly half the difference seen in MS10 ($\Delta \beta \approx -0.75$).  To find the difference in Hubble residuals between these two subsamples, we simulate the two subsamples with the derived parameters for the model color distributions.  We determine that the difference between the Hubble residuals when these two samples are combined is $0.02\pm0.003$ mags.  If the bias in $\beta$ was similar to that seen in MS10, the difference in Hubble residuals would be $0.036\pm0.005$ mags.  While this difference is statistically significant, it only composes a fraction of the difference in Hubble residuals dependent on host galaxy properties seen in most studies ($\sim0.07-0.08$ mag; \citealp{Ch13}).  More work must be done to further understand the difference in the values of $\beta$ found in MS10.


We also explore the significance of the relations between mass and Hubble residuals after we apply the BALT correction.  We find that following the conventional SALT2 approach, attributing residual scatter to luminosity, there is a difference in Hubble residuals of $0.075\pm0.014$ mag for SNe in high and low massive hosts.  With the BALT correction, we still find a difference of $ 0.062\pm0.016$ mag even though the reduction in $\chi^2$ from the BALT method is roughly $10\times$ the reduction from the host-galaxy correction ($\Delta \chi^2 \textrm{of} \sim5\%$).  We note though that for both SALT2 \citep{Ch13} and BALT, there is a remaining trend between color and Hubble residuals.  If we correct the distance modulus of each SN for this trend, the mass-Hubble residual effect is decreased by $\sim0.01$ mag.  Therefore we conclude that accounting for color variation may weaken the trend between Host galaxy properties and Hubble residuals, but this reduction alone is not large enough to explain the trend between Hubble residuals and host galaxy mass. 
\subsection{Evolution of Color-Luminosity Relation with Redshift}
\label{sec:evolution}

\cite{K09a} and \cite{Gu10} found that there is a non-negligible change in the color-luminosity relation with redshift in the SDSS and SNLS3 samples respectively.  We now explore whether the observed change of $\beta$ with redshift results from color variation.  
\begin{deluxetable*}{lcccc}
\tablecaption{$\beta$ evolution in Data and Simulations
\label{tab:evot}}
\tablehead{
\colhead{Sample} &
\colhead{Data} &
\colhead{Sim. Color (C11)} &
\colhead{Sim. Color (Smear)} &
\colhead{Sim. Lum (Guy10)}  \\
~&$\beta,~~~~\partial \beta / \partial z$&$\beta,~~~~\partial \beta / \partial z$&$\beta,~~~~\partial \beta / \partial z$&$\beta,~~~~\partial \beta / \partial z$ \\
}
\startdata

SNLS3 & $4.260 \pm 0.484,-2.378 \pm 0.847$ &  $3.950 \pm 0.109,-1.407 \pm 0.187$&  $2.741 \pm 0.313,1.551 \pm 0.617$&$  3.185 \pm 0.066,0.006 \pm 0.122$  \\
SDSS & $3.360 \pm 0.662,-2.287 \pm 3.093$&  $4.331 \pm 0.091,-4.809 \pm 0.446$&  $4.113 \pm 0.090,-4.888 \pm 0.430$& $ 3.282 \pm 0.059,-0.462 \pm 0.294$ \\
\enddata
\tablecomments{$\beta$ and $\partial \beta / \partial z$ are determined together given the assumption that the reduced chi-squared of the sample must be unity and residual scatter is attributed to the luminosity term.  In the real data samples, the SNLS3 sample has 240 SN\,Ia and the SDSS sample has 91 SN\,Ia.  In the simulation, each sample has 10,000 SNe\,Ia.}
\end{deluxetable*}


To understand the dependence of $\beta$ evolution on the source of scatter, we compare the values for recovered $\beta$ evolution from real data samples and simulations of different sources of scatter.   The results are shown in Table 3.  We determine $\beta$ and $\partial \beta / \partial z$ together, and we attribute the residual scatter to luminosity variation.  We find that the Guy10, luminosity variation model, predicts negligible $\beta$ evolution (for SNLS3: $\partial \beta / \partial z=0.01\pm0.12$), even though the amount is significant in the real data (for SNLS3: $\partial \beta / \partial z=-2.38\pm0.85$).  We simulate with two different color variation models: the color smear model discussed in Section 2 and the C11 model discussed in Section 3.  The difference between these two models is that the magnitude of color variation in the C11 model is higher at the blue end of the SN\,Ia spectral model, which is sampled at high-z.  We find that the C11 model better predicts $\beta$ evolution.  

We also note that when we allow a non-zero $\partial \beta / \partial z$ for the real data, $\beta=4.26\pm0.48$, which is near the Milky Way extinction value.  This implies that color appears to follow the Milky Way reddening law at low-z where scatter, noise and selection effects are weaker.  One other possible explanation \citep{Mar11} for $\beta$ evolution is that at higher redshifts, the color range decreases, and there is less leverage from the tails of the color distribution to help determine $\beta$.  However, we find that if we reduce the color range of the sample, the observed $\beta$ evolution becomes negligible.


\section{Discussion}
\label{sec:discuss}

 If residual scatter originates from color variation, then understanding the cause of color variation is paramount.  Much focus in the last few years has been placed on the relation between host galaxy properties and Hubble residuals.  Since host galaxy properties correlate with stretch and color,  we stress that a correction to the distance modulus after light curve corrections are done may not be the ideal method.  This approach is analogous to observing the correlation between Hubble residuals and color, finding a property like velocity that correlates with color, and removing the bias by finding a relation between Hubble residual and velocity.  A better approach for these scenarios may be to use the host galaxy or velocity information to inform the priors in which the color or stretch values are found.  
 
We have begun to explore how to incorporate Bayesian prior information into the SALT2 light curve fitter.  The BALT approach reduces the intrinsic dispersion in the sample to nearly null, which shows the promise of this approach.  Biases from introducing a Bayesian prior still need to be explored, though we reiterate that not including a prior is equivalent to using a flat prior.  There are currently fitters, like BayesN \citep{Mandel/etal:2011}, that include detailed Bayesian priors already, and these offer helpful guidance.  \cite{Mandel/etal:2011} also explores the consistency of infrared observations of SNe\,Ia with a MW Reddening law and finds they are consistent.  \cite{phillips11} reaches a similar conclusion for the majority of the SNe with normal ($E(B-V) < 0.3$) extinction values.

While we have shown how biases in $\beta$ and $w$ depend on the width and shape of the model color distribution, there are further complexities to an understanding of this distribution.  For our dust models, we assume that the peak of the model color distribution is at the $A_V=0$ limit.  It is possible that the peak is actually more red than this blue cutoff, as suggested from the analysis of \cite{2009ApJ...700..331H}.  We have also assumed that residual variation in color or luminosity is gaussian.  However, \cite{Fo11a} show that the velocity of SNe\,Ia is tied to color, and SNe\,Ia are not distributed evenly among high and low velocities.  Furthermore, we have not yet explored the consistency of infrared observations with the analysis done here.  


Finally, we mention that although the focus of this paper has been on the SALT2 fitter, the biases discussed in this paper will affect any light curve fitter that assumes the source of residual scatter among most SNe\,Ia is due to luminosity variation or ignores the effects of the underlying color distribution.   Another potentially fruitful path to characterizing intrinsic scatter is to reduce it through further sub-typing or discovery of additional SN parameters.
\section{Conclusions}
\label{sec:conclus}

In this paper, we have explained how fitting SN\,Ia distances depends on assumptions about the residual scatter of SN\,Ia.  We have also introduced a discussion of the biases due to ignorance of the model color distribution.  We show that the combination of residual scatter due to color and a realistic color distribution will bias $\beta$ by roughly $\sim1$ lower than its true value.  We find that a model in which color is solely due to Milky Way-like reddening along with residual color variation can explain the trend between Hubble residuals and color seen in the SNLS3, SDSS and Nearby data sets.  We also argue that an empirical-only analysis of light curve data contains multiple degenerate parameters, and further progress may stem from including astrophysical priors.  We have shown one method to include knowledge of the color model as a Bayesian prior, and how this approach significantly reduces the intrinsic dispersion in the sample.

Finally, the ultimate goal for SN\,Ia analysis is to measure cosmological parameters.  We find that misattributing the source of residual scatter can bias $w$ by as much as $4\%$.  This amount has been significantly underestimated in the past.  Further improvements in the determination of $w$ may come from a better understanding of the residual scatter of SN\,Ia and the true nature of SN\,Ia color.


\begin{thebibliography}{9}
\bibliographystyle{apj}

\bibitem[{{Burns} {et~al}\mbox{.}(2011){Burns}, {Stritzinger}, {Phillips},
  {Kattner}, {Persson}, {Madore}, {Freedman}, {Boldt}, {Campillay},
  {Contreras}, {Folatelli}, {Gonzalez}, {Krzeminski}, {Morrell}, {Salgado}, \&
  {Suntzeff}}]{SNooPy}
{Burns} C.~R. {et~al.}, 2011, \aj, 141, 19

\bibitem[{{Cardelli} {et~al.}(1989){Cardelli}, {Clayton}, \& {Mathis}}]{ccm89}
{Cardelli}, J.~A., {Clayton}, G.~C., \& {Mathis}, J.~S. 1989, \apj, 345, 245

\bibitem[Childress et al.(2013)]{Ch13} Childress, M., et al. 2013, \apj, 770, 107

\bibitem[{Chotard(2011)}]{Chotard_thesis}
Chotard, N. 2011, PhD thesis, University Claude Bernard Lyon, 1, Lyon, France

\bibitem[{Chotard {et~al.}(2011)}]{Chotard_2011}
Chotard, N. {et~al.} 2011, \aap, 529, L4

\bibitem[{{Conley} {et~al.}(2008)}]{2008ApJ...681..482C} 
{Conley}, A. {et~al.} 2008, \apj, 681, 482 


\bibitem[Conley et al.(2011)]{Conley_etal_2011} Conley, A., et al. 2011, ApJSS, 192, 1


\bibitem[{{Contreras} {et~al.}(2010)}]{2010AJ....139..519C} 
{Contreras}, C. {et~al.} 2010, \aj, 139, 519 

\bibitem[{Eisenstein {et~al.}(2005)}]{Eisen05}
Eisenstein, D. {et~al.} 2005, \apj, 633, 560

\bibitem[Foley \& Kasen(2011)]{Fo11a} Foley, R. and Kasen, D., 2011, \apj, 729, 55.

\bibitem[Foley et al.(2011)]{Fo11} Foley, R.~J., Sanders, 
N.~E., \& Kirshner, R.~P.\ 2011, \apj, 742, 89 

\bibitem[{{Gupta} {et~al.}(2011){Gupta}, {D'Andrea}, {Sako}, {Conroy}, {Smith},
  {Bassett}, {Frieman}, {Garnavich}, {Jha}, {Kessler}, {Lampeitl}, {Marriner},
  {Nichol}, \& {Schneider}}]{gupta2011}
{Gupta}, R.~R. {et~al.} 2011, \apj, 740, 92

\bibitem[{{Guy} {et~al.}(2007)}]{2007A&A...466...11G} 
{Guy}, J. {et~al.} 2007, \aap, 466, 11 


\bibitem[Guy et al.(2010)]{Gu10} Guy, J. et al., 2010, \aap, 523, A7.


\bibitem[{{Hicken} {et~al.}(2009{\natexlab{a}})}]{2009ApJ...700..331H} 
{Hicken}, M. {et~al.} 2009{\natexlab{a}},  \apj, 700, 331 

\bibitem[{{Hicken} {et~al.}(2009{\natexlab{b}})}]{2009ApJ...700.1097H} 
{Hicken}, M., {Wood-Vasey}, W.~M., {Blondin}, S., {Challis}, P., {Jha}, S., 
 {Kelly}, P.~L., {Rest}, A., \& {Kirshner}, R.~P.  2009{\natexlab{b}}, 
\apj,  700, 1097 

\bibitem[Holtzman et al.(2008)]{Ho08} Holtzman, J.A., et al. 2008, \aj, 136, 2306.


\bibitem[{{Jha} {et~al.}(2006)}]{2006AJ....131..527J} 
{Jha}, S. {et~al.} 2006, \aj, 131, 527 



\bibitem[{Jha, Riess \& Kirshner}(2007)]{Jh07} Jha, S., Riess, A. G., and Kirshner, R. P., 2007, \aj, 659, 122.



\bibitem[{{Kelly} {et~al.}(2010)}]{2010ApJ...715..743K} 
{Kelly}, P.~L., {Hicken}, M., {Burke}, D.~L., {Mandel}, K.~S., \& 
{Kirshner},  R.~P.  2010, \apj, 715, 743 


 \bibitem[Kessler et al.(2009a)]{K09a} Kessler, R., et al. 2009, \pasp, 121, 1028.
 \bibitem[Kessler et al.(2009b)]{K09b} Kessler, R. et al., 2009,  \apjs, 185, 32.

\bibitem[{{Kessler} {et~al.}(2013){Kessler}, {Guy}, {Marriner}, {Betoule},
  {Brinkmann}, {Cinabro}, {El-Hage}, {Frieman}, {Jha}, {Mosher}, \&
  {Schneider}}]{kessler_13_var} Kessler, R. et al., 2013, \apj, 764, 48

\bibitem[Kim et al. (2013)]{kimgp} Kim, A.G., et al. 2013, \apj, 766, 84.

\bibitem[{{Komatsu} {et~al.}(2009)}]{Komatsu08}
{Komatsu}, E. {et~al.} 2009, \apjs, 180, 330

\bibitem[Lira(1995)]{Lira} Lira, P.\ 1995, Masters Thesis, Universidad de Chile

\bibitem[{{Mandel}, {Narayan} \& {Kirshner}(2011){Mandel}, {Narayan}, \&
  {Kirshner}}]{Mandel/etal:2011}
{Mandel} K.~S., {Narayan} G., {Kirshner} R.~P., 2011, \apj, 731, 120



\bibitem[{{Marriner} {et~al.}(2011){Marriner}, {Bernstein}, {Kessler},
  {Lampeitl}, {Miquel}, {Mosher}, {Nichol}, {Sako}, {Schneider}, \&
  {Smith}}]{Mar11}
{Marriner}, J., {Bernstein}, J.~P., {Kessler}, R., {et~al.} 2011, \apj, 740, 72



\bibitem[{Perlmutter {et~al.}(1999)}]{Saul99}
Perlmutter, S. {et~al.} 1999, \apj, 517, 565

\bibitem[{{Phillips}(1993)}]{phillips93}
{Phillips}, M.~M. 1993, \apjl, 413, L105

\bibitem[{Phillips {et~al.}(1999)}]{Phillips_99}
Phillips, M.~M. {et~al.} 1999, \aj, 118, 1766

\bibitem[Phillips(2011)]{phillips11} Phillips, M.~M.\ 2011, PASA, in
  press, arXiv:1111.4463

\bibitem[{{Riess}, {Press}, \& {Kirshner}(1996)}]{riess96}
---. 1996, \apj, 473, 88

\bibitem[{Riess {et~al.}(1998)}]{Riess98}
Riess, A. {et~al.} 1998, \aj, 116, 1009









\bibitem[{{Schlegel}, {Finkbeiner}, \&  
{Davis}(1998)}]{1998ApJ...500..525S} 
{Schlegel}, D.~J., {Finkbeiner}, D.~P., \& {Davis}, M.  1998, \apj, 500, 
525 



\bibitem[{{Sullivan} {et~al.}(2010)}]{2010MNRAS.406..782S} 
{Sullivan}, M. {et~al.} 2010, \mnras, 406, 782 

\bibitem[{{Sullivan} {et~al.}(2006)}]{2006ApJ...648..868S} 
{Sullivan}, M. {et~al.} 2006, \apj, 648, 868 

\bibitem[{{Suzuki} {et~al.}(2012){Suzuki}, {Rubin} \& {Supernova Cosmology Project}}]{suzuki_2012_aa} {Suzuki}, N., {Rubin}, D., {Lidman}, C., {et~al.} 2012, \apj, 746, 85

\bibitem[{{Wang} {et~al.}(2009){Wang}, {Filippenko}, {Ganeshalingam}, {Li},
  {Silverman}, {Wang}, {Chornock}, {Foley}, {Gates}, {Macomber}, {Serduke},
  {Steele}, \& {Wong}}]{2009ApJ...699L.139W}
{Wang}, X., {et~al.} 2009, \apjl, 699, L139




\end{thebibliography}

\acknowledgments

:
We thank Rick Kessler for many useful conversations.

\appendix
\section{SALT2mu}
To determine the nuisance functions $M_0$, $\alpha$ and $\beta$, the SN sample is divided into $>5$ equally sized redshift bins and $M_0,\alpha$ and $\beta$ are found in each bin to minimize the distance modulus scatter relative to a trial cosmology.  This process is done by the routine SALT2mu (M11), and for each bin, the $\chi^2$ is minimized where
 \begin{multline}
\chi^2 = \displaystyle\sum_{n=1}^N 
[\mu_{\textrm{n}}
-\mu_{\textrm{t}}(z_n,\Omega,w)]^2/(\sigma_n^2+\sigma_{\rm{r}}^2) = 
\displaystyle\sum_{n=1}^N 
[m_{Bn}-M_0+\alpha x_{1n}-\beta c_{n} 
-\mu_{\textrm{tr}}(z_n,\Omega,w)]^2/(\sigma_n^2+\sigma_{\rm{r}}^2).  
\label{eqn:s2mchi2}
\end{multline}
For the $n^{\rm{th}}$ SN, $\sigma_n$ is the error from the light curve fit, $\sigma_{\rm{r}}$ is the total residual scatter and $\mu_\textrm{t}$ is a trial cosmology given the matter density of the universe $\Omega$ and the equation of state parameter $w$.  Since $M_0$ is allowed to vary and it is degenerate with the cosmology, the determination of $\alpha$ and $\beta$ will be independent of the cosmology.  This is shown in both M11 and K13. Given the best fit $\alpha$ and $\beta$, and that $\mu_{\textrm{t}}$ is the best fit cosmology, the numerator of the expressions in Eqn.~\ref{eqn:s2mchi2} represents the `Hubble residuals'.

\section{Color Distribution}

Assuming the true color population of the SNe is a gaussian, the distribution of the observed colors may be expressed as:
\begin{multline}
 e^{ [-(c_{\textrm{obs}})^2/2(\sigma_{c_{\textrm{obs}}}^2)]}=    e^{ [-(c_{\textrm{mod}})^2/2(\sigma_{c_{\textrm{mod}}}^2)]} \ast    e^{ [-(c_n)^2/2(\sigma_{c_n}^2)]} \ast    e^{ [-(c_r)^2/2(\sigma_{c_r}^2)]}=    e^{ [-(c)^2/2(\sigma_{c_{\textrm{mod}}}^2+\sigma_{c_n}^2+\sigma_{c_r}^2)]}.
\label{eqn:specf2}
\end{multline}

Therefore, we find that 
\begin{equation}
\sigma^2_{c_{\textrm{obs}}}=\sigma^2_{c_{\textrm{mod}}}+\sigma^2_{c_n}+\sigma^2_{c_r}.
\end{equation}
To find the relation between values of $\beta$ for different models of color, one must take the square of the derivative of the distance modulus (Eq. 1) with respect to $c$.
So that the distribution of $c_{\textrm{obs}}$ from the SNANA Monte-Carlo (MC) simulations of the SDSS and SNLS3 samples match the data, K13 retroactively derives the true color distribution of $c_{\textrm{mod}}$ for both the SNLS3 and SDSS surveys.  K13 finds that an asymmetric gaussian is needed to describe both the input stretch and color distribution.  Here we express the function for color, and note that stretch can be defined in the same manner: 
\begin{eqnarray}
   e^{ [-(c_{\textrm{mod}} - \bar{c_{\textrm{mod}}})^2/2{\sigma_{c_{\textrm{mod}}}}^2] } & ~~~~ & c_{\textrm{mod}} < \bar{c}_{\textrm{mod}} \\
   e^{ [-(c_{\textrm{mod}} - \bar{c_{\textrm{mod}}})^2/2{\sigma_{cmod+}}^2]  } & ~~~~ & c_{\textrm{mod}} > \bar{c}_{\textrm{mod}}. 
\label{eqn:specf}
\end{eqnarray}
In Eq.~\ref{eqn:specf}, $\bar{c_{\textrm{mod}}}$ is the peak value of the distribution, $\sigma_{cmod+}$ is the standard deviation of the colors redder than the mean and $\sigma_{c_{\textrm{mod}}}$ is the standard deviation of the colors bluer than the mean.  

\end{document}